\documentclass[aps,prb,twocolumn,groupedaddress,showpacs,showkeys]{revtex4}

\usepackage{graphicx}

\begin{document}


\title{Phase diagram of magnetic configurations for magnetic nanodots of circular and elliptical shape with perpendicular anisotropy}



\title{\bf Properties of magnetic nanodots with perpendicular anisotropy}%

\author{E. R. P. Novais$^{1}$}
\author{P. Landeros$^{2}$}
\author{A. G. S. Barbosa$^{3}$}
\author{M. D. Martins$^{3}$}
\author{F. Garcia$^{4}$}
\author{A. P. Guimar\~aes$^{1}$}

\affiliation{$^{1}$Centro Brasileiro de Pesquisas F\'{\i}sicas, 22290-180,  Rio de Janeiro, RJ, Brazil }%

\affiliation{$^{2}$Dept. de F\'{\i}sica, Universidad T\'ecnica Federico Santa Maria, Valparaiso, Chile},

\affiliation{$^{3}$Centro de Desenvolvimento da Tecnologia Nuclear, 31270-901, Belo Horizonte, MG, Brazil}

\affiliation{$^{4}$Laborat\'orio Nacional de Luz S\'{\i}ncrotron, 13083-970, Campinas, SP, Brazil}


\date{\today}

\begin{abstract}
Nanodots with magnetic vortices have many potential applications, such as magnetic memories (VRAMs) and spin transfer nano-oscillators (STNOs). Adding a perpendicular anisotropy term to the magnetic energy of the nanodot it becomes possible to tune the vortex core properties. This can be obtained, e.g., in Co nanodots by varying the thickness of the Co layer in a Co/Pt stack. Here we discuss the spin configuration of circular and elliptical nanodots for different perpendicular anisotropies; we show for nanodisks that micromagnetic simulations and analytical results agree. Increasing the perpendicular anisotropy, the vortex core radii increase, the phase diagrams are modified and new configurations appear; the knowledge of these phase diagrams is relevant for the choice of optimum nanodot dimensions for applications. MFM measurements on Co/Pt multilayers confirm the trend of the vortex core diameters with varying Co layer thicknesses.
\end{abstract}



\maketitle


\section{Introduction}

Nanoscopic and mesoscopic magnetic structures have attracted the interest of many workers in
recent years in view of their very interesting physical properties and for their potential applications. Quasi-twodimensional magnetic nanodots made of soft magnetic materials, such as permalloy, may present for their lowest energy state, several magnetic configurations: (i) quasi-uniform in-plane state, (ii) quasi-uniform out-of-plane state, (iii) magnetic vortices or swirls \cite{Guimaraes:2009}.

Magnetic vortices are structures where the magnetic moments are tangential to concentric circles. The center of the vortex has a singularity (the vortex core) where the magnetization points out of the plane, with a radius, in the thin dot limit \cite{Hubert:1999}, of the order of the exchange length of the material
$l_{\mathrm ex}=\sqrt{2A/\mu_0 M_s^2}$, where $A$ is the exchange stiffness constant and $M_s$ is the saturation magnetization \cite{Chien:2007, Guslienko:2008}. With the parameters used in the present work, for permalloy $l_{\mathrm{ex}}=5.3$ nm and for cobalt, $l_{\mathrm ex}$ = 4.93 nm.

Magnetic vortices have been observed by many experimental techniques, such as magnetic force microscopy \cite{Shinjo:2000}, X-ray microscopy \cite{Choe:2004}, or inferred from hysteresis curves \cite{Cowburn:1999}; they also result from theoretical modeling \cite{Albuquerque:2002, Scholz:2003, Jubert:2004, Landeros:2005, Kravchuk:2007, Landeros:2008}.

The proposed applications of magnetic nanodots include their use in patterned magnetic recording media \cite{Thomson:2008}, as elements in magneto-resistance random access memories (MRAM's) \cite{Kim:2008, Bohlens:2008}, spin transfer nano-oscillators (STNO's) \cite{Berkov:2009, Lehndorff:2009, Houssameddine:2007} and nanoscopic agents for cancer treatment \cite{Kim:2010}.

The different magnetic configurations, and consequently the large variation in the magnetic properties observed in nanodots as a function of dimensions, underline the interest in the study of diagrams (phase diagrams) mapping the parameter space where a given magnetic behavior is to be expected.


Experimental studies have been used to obtain the phase diagram for magnetic disks: Ross et al. \cite{Ross:2002} derived the phases from hysteresis curves, Chung et al. \cite{Chung:2010} from SEMPA measurements.
Metlov and Guslienko \cite{Metlov:2002} obtained a phase diagram with regions of in-plane magnetization, perpendicular magnetization and vortex structure; the equilibrium magnetic configuration obtained by micromagnetic simulation showed general agreement with this diagram\cite{Scholz:2003, Chung:2010}. Another simulation study, this time using a scaling approach, was made for circular and elliptical nanodots \cite{Zhang:2008}. They have found in the phase diagram for ellipses a double vortex arrangement for dots with semi-axis $a$ larger than 150 nm. Their simulations were made for core-free ellipses, a choice that might displace the phase boundaries by a significant 35\%.

We have recently shown theoretically and experimentally\cite{Garcia:2010}, that using Co/Pt multilayers it is possible to tailor the vortex core diameter by playing with the perpendicular anisotropy originated at the Co-Pt interface. When one increases the perpendicular anisotropy acting on a magnetic nanodot, e.g., reducing the Co layer thickness, the vortex core diameter increases, and eventually another vortex state appears, which is characterized by an out-of-plane magnetization component at the dot rim. The increase in perpendicular anisotropy has an effect that is somehow equivalent to a reduction in the relative importance of the in-plane shape anisotropy.

The main goal of the present work is to study how the phase diagram of magnetic nanodots is modified by the presence of perpendicular anisotropy ($K_z$). However, we have first obtained the phase diagram for nanodots with $K_z=0$, with circular and elliptical shapes. This has been done for two main reasons, first to illustrate and validate our methodology, which will be used in sequence in this paper, and to verify the effect of the magnetostatic energy (responsible for shape anisotropy) on the ground state magnetic arrangement.

We have obtained phase diagrams by micromagnetic simulation and analytically, and they are in agreement. We have also indicated experimentally, using Magnetic Force Microscopy (MFM), that the vortex core diameter shows the same trend as predicted by the theory. The paper is organized as follows: in section II we discuss magnetic configurations of the disks, which include IIA) micromagnetic simulations, IIB) analytical description, IIC) experimental results. In Section III we present the results for the ellipses: IIIA) micromagnetic simulations. In Section IV we present a brief discussion, a summary of the main results with the conclusions.

\section{Results for disks}


\subsection{Disks: Micromagnetic Simulations}

For the simulations we used the OOMMF package (a free software available from NIST\footnote{Available from  http://math.nist.gov/oommf/}), using the parameters for bulk permalloy, to allow a comparison with the literature (exchange stiffness constant $A=1.3\times10^{-11}$ J/m, saturation magnetization $M_s=860\times 10^3$ A/m). We have neglected the in-plane anisotropy; however, we have also simulated magnetic systems exhibiting a perpendicular anisotropy. An application of such simulations is the description of the behavior of the Co/Pt multilayers. To make the present results of more general use, they have been given in terms of normalized parameters, using the exchange length $l_{\mathrm ex}$.

For some dimensions of the nanodots, the simulation may converge to a configuration that does not
correspond to the absolute energy minimum. Therefore, the simulations made with parameters near the
boundary regions between different configurations had to be made by initially imposing different magnetic configurations, and after the convergence, comparing the resulting energies to determine the spin arrangement corresponding to the absolute energy minimum. The boundary lines between the different phases were obtained from the intersection of the curves of total energy for the different states \cite{Novais:2009}.

Effects of discretization are inherent in the methodology used here (see \cite{Jubert:2004}). For this reason we have studied the effect of the cell size (for most of our simulations $5\times 5 \times 5$ nm$^3$). We have found that the position of the boundaries of our phase diagrams change very little for different cell sizes; these effects are even less important for the configuration with perpendicular magnetization.

\begin{figure}
\includegraphics[width=\columnwidth]{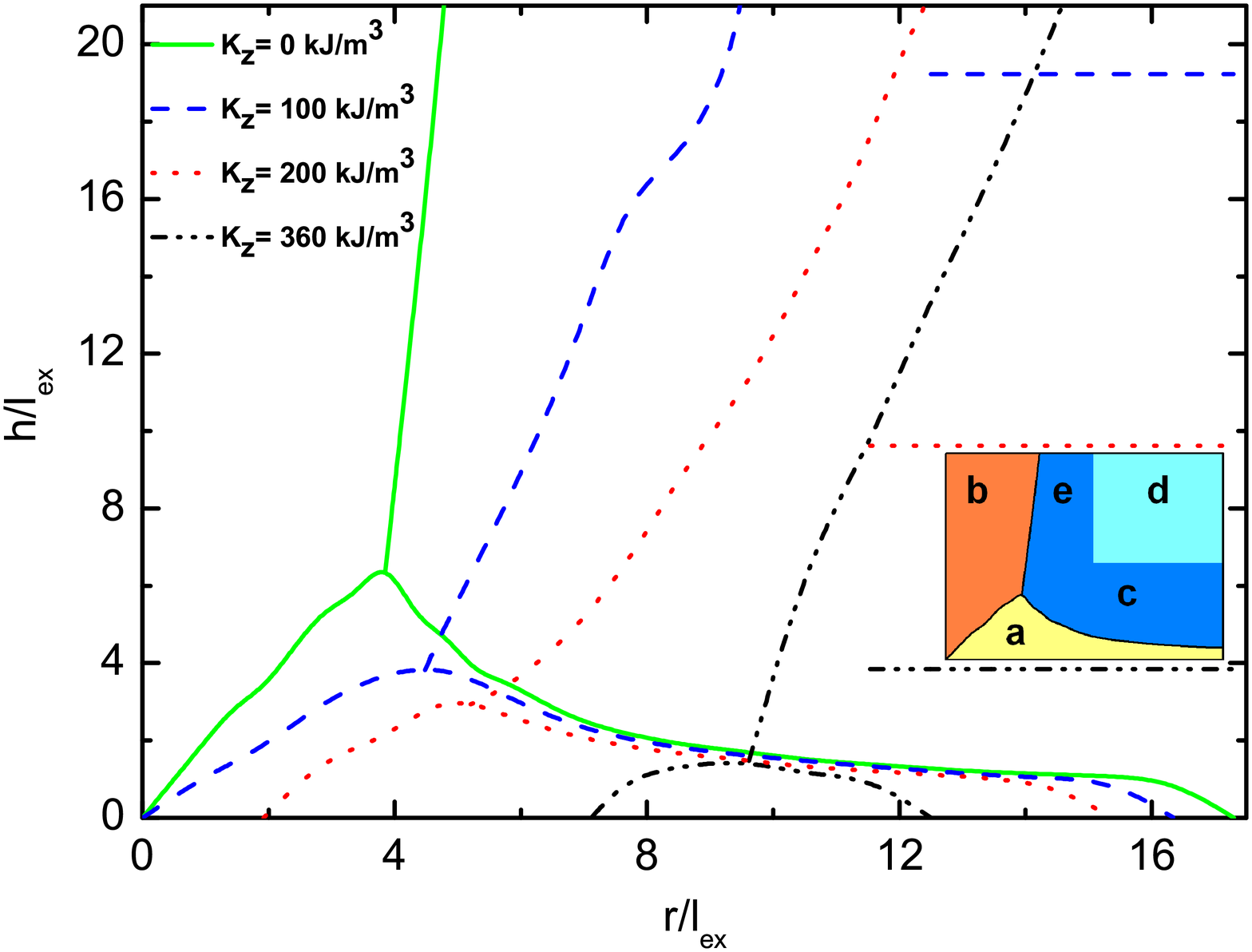}
\caption{\label{fig:DiskPhaseDiagramAnis}(Color online) Phase diagrams for circular nanodots as a function of reduced height $h/l_{\mathrm ex}$ and reduced radius $r/l_{\mathrm ex}$, drawn from the minimum energy computed with micromagnetic simulation for different values of the perpendicular anisotropy: $K_z=0$, $100 \times 10^3$ J/m$^3$, $200 \times 10^3$ J/m$^3$, $360 \times 10^3$ J/m$^3$. The different magnetic configurations are labeled in the inset  showing the $K_z=0$ phase diagram: a) single domain parallel to the plane, b) single domain perpendicular to the plane, c) vortex, d) configuration given by Fig. \ref{fig:Configurations}d, and e) vortex with perpendicular component (Fig. \ref{fig:Configurations}e). }
\end{figure}

\begin{figure}
\includegraphics[width=\columnwidth]{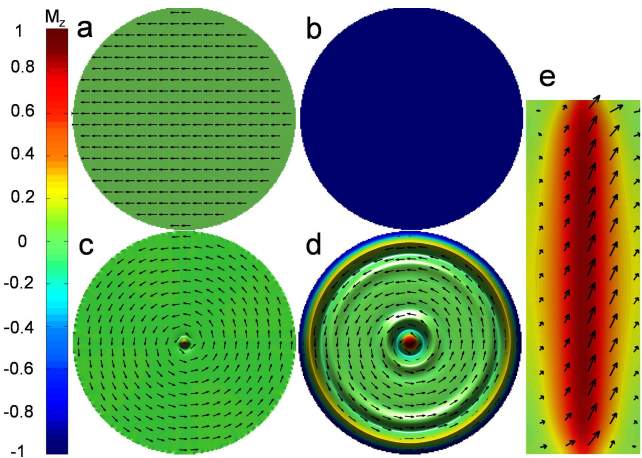}
\caption{\label{fig:Configurations}(Color online) Magnetization configurations for nanodisks:  a) quasi-uniform in-plane magnetization; b) quasi-uniform perpendicular magnetization; c) magnetic vortex; d) disk with parameters in the region above the red continuous line (or the blue dotted line) in Fig. \ref{fig:MicromagSimulRadius} (d=400 nm and anisotropy $K_z=375 \times 10^3$ J/m$^3$); e) lateral view showing the longitudinal section of an elongated nanodisk as found in the region of the phase diagram where the vortex acquires a perpendicular magnetization component (see region e in Fig. \ref{fig:DiskPhaseDiagramAnis}).}
\end{figure}

The phase diagram for magnetic nanodisks obtained from the computed energies for the different magnetization configurations and for different perpendicular anisotropy values ($K_z$) is shown in Fig. \ref{fig:DiskPhaseDiagramAnis}.
For $K_z=0$ the phase diagram agrees with those of references \cite{Albuquerque:2002, Ross:2002, Scholz:2003, Jubert:2004, Landeros:2005, Kravchuk:2007, Landeros:2008, Metlov:2002, Zhang:2008}. It shows three regions, depending on the aspect ratio of the disks. The corresponding magnetic configurations are shown in Fig. \ref{fig:Configurations}. For very thin disks, for a wide range of disk radii, the shape anisotropy favors a quasi-uniform in-plane state; on the other hand, for thicker disks and approximately $r < 4$ $l_{ex}$, a quasi-uniform out-of-plane state is observed, that is easy to understand, since in this region nanodots cannot be taken as approximately twodimensional disks. Finally, for $r > 4$ $l_{ex}$ and $h > 4$ $l_{ex}$ we observe different magnetic vortex states as the ground states. One should note that, in the vortex state region in the graph, the magnetization shows an increasing out-of-plane component as the disk thickness increases; this can be seen in Fig. \ref{fig:Configurations}e. Note that the vortex core diameter varies along the length of the cylinder, reaching a maximum at half the height. Also, we observe the formation of a "mixed" state (e in Fig. \ref{fig:DiskPhaseDiagramAnis}), where the magnetization shows vortex domains at the dot ends, and out of plane magnetization near half height. An axial section of a nanodisk with larger $h$ (Fig. \ref{fig:Configurations}d) shows that the vortex acquires a perpendicular magnetization component.

When we include a perpendicular anisotropy term, the phase diagram of the disks is modified, as can be seen in Fig. \ref{fig:DiskPhaseDiagramAnis}. As expected, the region corresponding to magnetization perpendicular to the plane (region b, shown in the inset) is increased as $K_z$ is increased, displacing to larger radii the boundary line between the quasi-uniform perpendicular moment state and the vortex state. Furthermore, the region for in-plane magnetization (region a) is reduced. In the case of the simulation with the highest perpendicular anisotropy shown in the figure ($K_z=360 \times 10^3$ J/m$^3$), the in-plane magnetization region is limited to a narrow range between 7 and 12 $l_{ex}$, for very thin disks. The tendency of an out-of-plane magnetization in the vortex region is not observed in the cases of nonzero perpendicular anisotropy.
For higher anisotropies another more complex configuration appears with an out-of-plane magnetization at the disk rim (region d), as observed experimentally in ref. \cite{Garcia:2010}.

Increasing the perpendicular anisotropy increases the vortex core radius, and eventually leads to a more complex spin structure, with the magnetization at the disk rim pointing down (Fig. \ref{fig:Configurations}d). Further increase in the perpendicular anisotropy leads to a uniform perpendicular magnetization, as shown in Fig. \ref{fig:Configurations}b.

\begin{figure}
\includegraphics[width=\columnwidth]{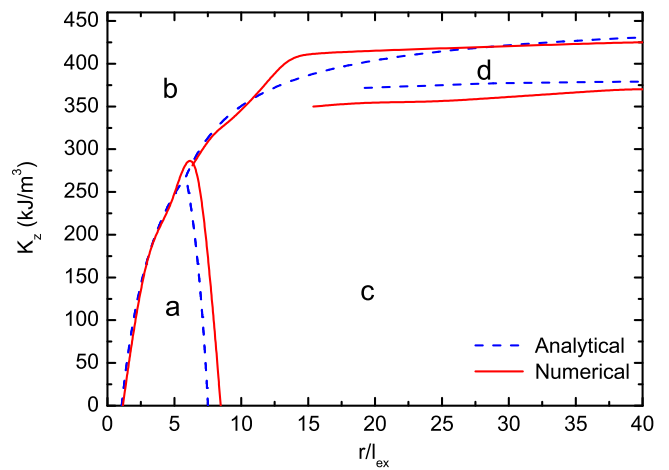}
\caption{\label{fig:MicromagSimulRadius}(Color online) Phase diagram for a 10 nm-thick magnetic disk, as a function of the perpendicular anisotropy $K_z$ and reduced radius $r/l_{ex}$ for micromagnetic simulation (red continuous line) and analytical computation (blue dotted line). The spin configurations are: a) single domain parallel to the plane, b) single domain perpendicular to the plane, c) vortex, d) configuration given by Fig. \ref{fig:Configurations}d. The lines on the righthand side  (red continuous=simulation, blue dotted=analytic) limit the region above which the spin structure is given by d.}
\end{figure}

 The dependence of the magnetic structure of the disks with the value of the perpendicular anisotropy is exhibited in Fig. \ref{fig:MicromagSimulRadius}. Here we have fixed the thickness of the nanodot (10 nm) and obtained the magnetic phase diagram in the plane of perpendicular anisotropy versus reduced radius of the dot derived both by micromagnetic simulation (red continuous line), and obtained analytically (blue dotted line). The agreement between the two methods is very good.

\subsection{Disks: Analytical Method}

In order to describe the configurations of magnetic nanodisks we have developed a simple model for the magnetic vortex state with out-of-plane magnetization at the dot rim. We take into account volume ($K_v$) and perpendicular anisotropy ($K_z$), as well as dipolar and exchange energy contributions.

The energy of the magnetic states with in-plane (IP) and out-of-plane (OP) uniform magnetization can be written as \cite{Jubert:2004, Landeros:2008}:

\begin{equation}
E^{IP}=\frac{\mu_0 M_s^2}{4}\pi r^2h[1-N_z(r, h)]
\end{equation}

\begin{equation}
E^{OP}=\frac{\mu_0 M_s^2}{2}\pi r^2h\left[N_z(r, h)+\frac{2(K_v-K_z)}{\mu_0 M_s^2}\right]
\end{equation}
where $N_z$ is the demagnetizing factor \cite{Landeros:2008}.

$N_z$ is given by \cite{Jubert:2004}:

\begin{equation}
N_z=1 +  \frac{8r}{3 \pi h} - F_{21}\left[ -\frac{1}{2}, \frac{1}{2}, 2, \frac{4 r^2}{h^2} \right]
\end{equation}
where $J_1(x)$ is the first order Bessel function and $F_{21}(a, b, c, x)$ is the hypergeometric function.

The vortex states can be generally described in terms of the magnetization $M_z(\rho) = M_s m_z(\rho)$, and it can be shown \cite{Jubert:2004} that the relevant energy terms can be written as

\begin{equation}
E_d=\pi\mu_0 M_s^2 \int_0^\infty dq \left( \int_0^r \rho J_0(q\rho)m_z(\rho)d\rho \right)^2 \left(1- e^{-qh}\right)
\end{equation}

\begin{equation}
E_{ex}=2\pi A h \int_0^r \left[\frac{1-m_z^2(\rho)}{\rho^2} + \frac{1}{1-m_z^2(\rho)}  \left( \frac{\partial m_z(\rho)}{\partial \rho} \right)^2 \right]\rho d\rho
\end{equation}

\begin{equation}
E_K=2\pi h \left(K_v- K_z \right)\int_0^r m_z^2(\rho)\rho d\rho
\end{equation}
where $J_0(x)$ are Bessel functions. For the vortex states we consider the following {\it ansatz}:

\begin{equation}
m_z(\rho)= \left\{
\begin{array}{c c}
                       (1-\rho^2/b^2),         & 0<\rho<b \\
               0,                      & b<\rho<r-c \\
                       -g(1-(r-\rho)^2/c^2)^4, & r-c<\rho<r \\
\end{array} \right.
\end{equation}
where $b$ is a parameter related to the core radius \cite{Landeros:2008}, $c$ is related to the size of the out-of-plane magnetization at the rim of the dot, and $g$ ($0< g <1$) is used to describe the magnetization at the rim; for an usual vortex $g = 0$. With the above magnetization we perform a numerical evaluation of the total energy with minimization of the adjustable parameters $b$, $c$ and $g$.

This theoretical description has allowed the determination of the phase diagram as a function of perpendicular anisotropy, as well as the boundaries of the region of the diagram where the magnetic nanodisks exhibit perpendicular magnetization at the rim (Fig. \ref{fig:MicromagSimulRadius}).

\begin{figure}
\includegraphics[width=\columnwidth]{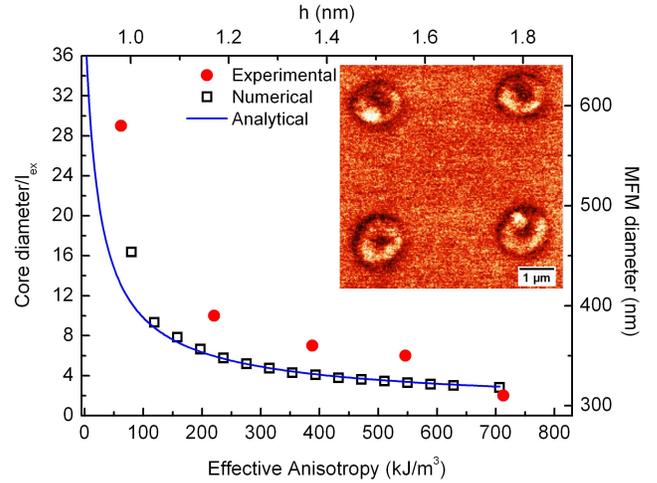}
\caption{\label{fig:CoreDiameterVsK}(Color online) Dependence of the magnetic vortex core diameter with the perpendicular anisotropy. The black squares are computed with micromagnetic simulation including  $K_z$; the circles (red) are experimental values obtained by MFM (righthand scale) of Co/Pt disks and the continuous (blue) line is obtained with the analytical model. Inset: MFM image of 1 $\mu$m diameter Co/Pt nanodisks showing the vortex cores.}
\end{figure}

The vortex core radius ($r_c$) can be defined as the value at which $m_z = 0.5$, and then $r_c=(1-2^{-1/4})b$, where $b$ is obtained by minimization of the energy. Using the magnetic parameters \cite{Garcia:2010} for bulk Co, (the exchange length is $l_{\mathrm ex}$ = 4.93 nm), we obtain the core size in the presence of perpendicular anisotropy, as shown in Fig. \ref{fig:CoreDiameterVsK}, in good agreement with Fig. 1 of Garcia et al. \cite{Garcia:2010}.

\begin{figure}
\includegraphics[width=\columnwidth]{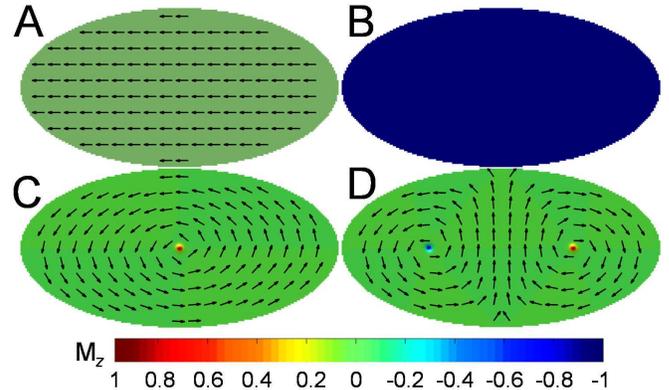}
\caption{\label{fig:EllipseConfigurations}(Color online) Magnetic configurations of elliptic nanodots that appear in the phase diagram of Fig. \ref{fig:EllipsePhaseDiagram}: A) in-plane quasi-uniform magnetization, B) out-of-plane quasi-uniform magnetization, C) single vortex configuration, D) double vortex configuration.}
\end{figure}

\vspace{5mm}

\subsection{Disks: Experimental}

 The samples were produced by magnetron sputtering deposition, by means of e-beam lithography on SiO$_2$/Si(100) wafers. The samples presented the same layer structure ([Co$_h$/Pt$_2$]$_6$/Pt$_6$) and were distinguished by the Co layer thickness ($h$ = 0.6, 0.8, 1.6, 2.0 nm) in the stack. We have chosen these thicknesses regarding a perpendicular to in-plane magnetic anisotropy transition observed when $h$ is increased from 0.4 to 0.8 nm.
Each sample contained arrays of 1 $\mu$m and 2 $\mu$m-diameter disks. For better comprehension of the results, a continuous film sample was produced along with each of the structured samples by placing a resist-free wafer on the side of the lithographed sample in the sputtering chamber. The quality of the lithography and deposition process has been verified by field emission gun scanning electron microscopy, Dektak profilometry and Rutherford Backscattering Spectroscopy.

For the MFM measurements we used an NTEGRA Aura MFM scanning probe
microscope (NT-MDT Co.) with a commercial MFM tip (NSG01 type, CoCr magnetic
coating, NT-MDT Co.) magnetized along the tip axis in the field of a
permanent magnet. The MFM images were acquired in the tapping mode at room temperature.
In order to avoid instrumental artifacts in the determination of vortex
core size from the MFM image we kept the lift height constant for all
the measured samples. Although MFM is not the most suitable technique
to determine quantitatively the vortex core diameter, we expected to obtain
the trend of the vortex diameter with the thickness of the Co layer
(Fig. \ref{fig:CoreDiameterVsK}).

We have determined experimentally the trend toward increasing vortex core diameter in the Co/Pt multilayers, as the perpendicular anisotropy acting on the nanodots is increased. The MFM measurements made on the Co disks to study this effect, however, do not allow the accurate determination of the vortex core diameter. They only allow the observation of this increasing trend, as shown in Fig. \ref{fig:CoreDiameterVsK}. In the figure, the vortex core diameters are plotted as a function of Co layer thickness or effective anisotropy ($K_{eff}=K_v - K_z$); in the analytical curve $K_v=0$. Fig. \ref{fig:CoreDiameterVsK} also shows the vortex core diameters obtained analytically and shows their agreement with the micromagnetic simulation results.

\begin{figure}
\includegraphics[width=\columnwidth]{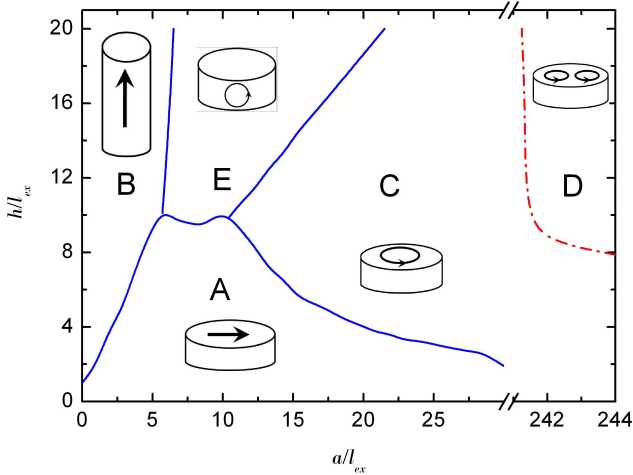}
\caption{\label{fig:EllipsePhaseDiagram}(Color online) Phase diagram for elliptical nanodots without perpendicular anisotropy ($K_z=0$), as a function of reduced height $h/l_{\mathrm ex}$ and reduced semi-axis $a/l_{\mathrm ex}$, drawn from the minimum energy computed with micromagnetic simulation. The letters correspond to regions with different spin configurations: A) in-plane quasi-uniform magnetization; B) perpendicular quasi-uniform magnetization; C) in-plane vortex; D) double in-plane vortex; E) lateral vortex. The ellipses in every case have semi-axes in the ratio $a/b=2$.}
\end{figure}

\section{Results for Ellipses}

\subsection{Ellipses: Micromagnetic Simulations}

Following the same methodology, we have also obtained the phase diagram of elliptical nanodots with $K_z=0$; we have simulated ellipses where the major semi-axis ($a$) is twice the minor one (b), i.e., the ellipses in every case have $a/b=2$. As we can see from Fig. \ref{fig:EllipsePhaseDiagram}, in this case the diagram is richer than that of the disks. The first observation is that the vortex state only occurs for dimensions that are larger than in the case of the disks, i.e., for approximately $a > 10$ $l_{ex}$. This is so because the eccentricity introduces an uniaxial shape anisotropy along the major axis, which favors a quasi-uniform in-plane magnetization.
Another state observed is the region in the figure which corresponds to one lateral vortex, that also appears in some simulations for cylinders \cite{Ha:2003}; to the best of our knowledge, this had not been observed for ellipses. A very interesting phase of this diagram occurs for $a > 240$ $l_{ex}$ and $h > 8$ $l_{ex}$, where two vortices appear. This configuration has been observed experimentally by several authors e.g., \cite{Schneider:2003, Buchanan:2005, Okuno:2004, Vavassori:2004}.

\begin{figure}
\includegraphics[width=\columnwidth]{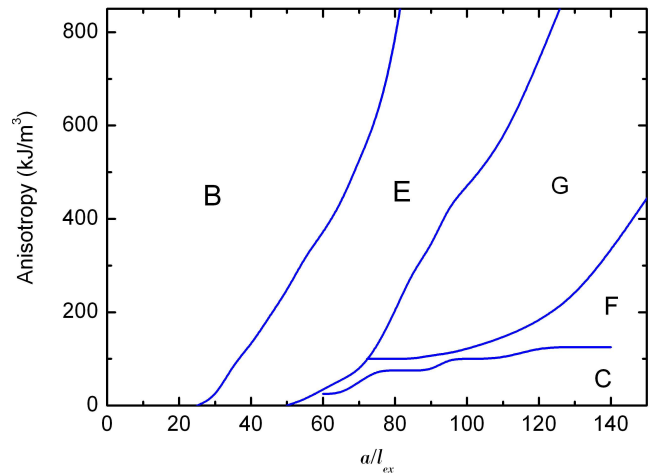}
\caption{\label{fig:EllipseDiagramKxa} Phase diagram for elliptical dots with thickness $h=50$ nm: diagram of perpendicular anisotropy $K_z$ versus reduced major semi-axis $a/l_{ex}$. The letters correspond to regions with different spin configurations:  B) perpendicular quasi-uniform magnetization, C) in-plane vortex, E) lateral vortex, F) modified in-plane vortex, and G) double lateral vortex (see Fig. \ref{fig:EllipseLateral}).}
\end{figure}

\begin{figure}
\includegraphics[width=\columnwidth]{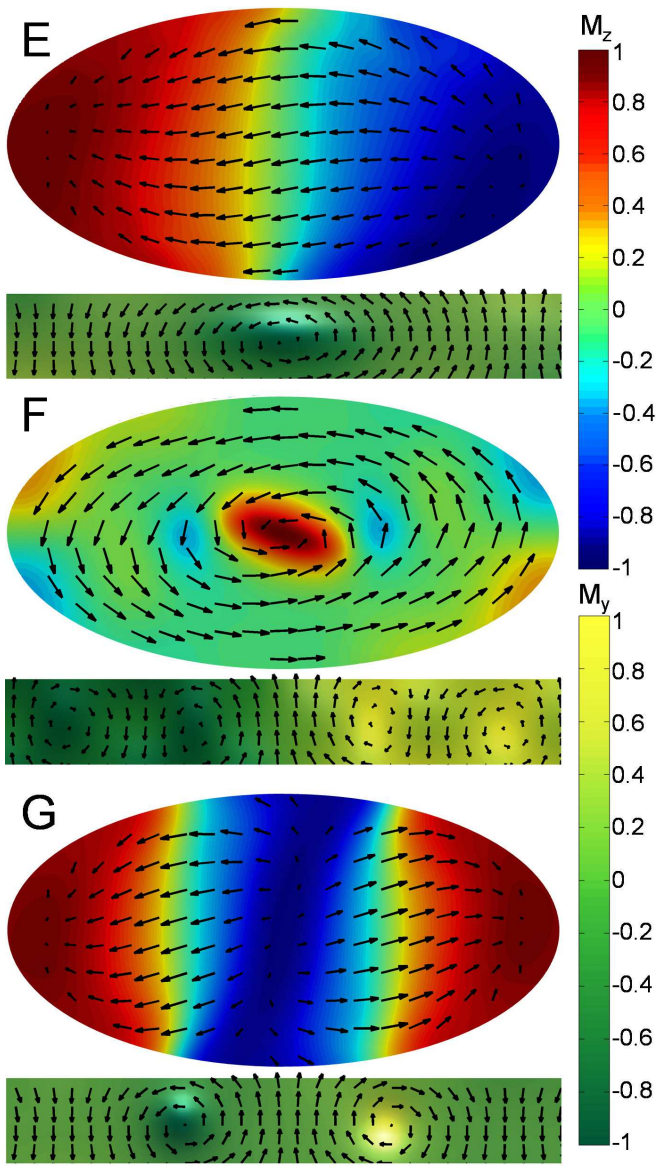}
\caption{\label{fig:EllipseLateral}(Color online) Magnetic configurations of elliptic nanodots, showing the plane of the dots and the cross sections with lateral vortices: E) "two-domain" configuration with single lateral vortex; F) modified in-plane vortex and G) "three domain" configuration with two lateral vortices of opposite polarization. Configuration E occurs in the phase diagram for $K_z=0$ (Fig. \ref{fig:EllipsePhaseDiagram}), and configurations F and G appear in the phase diagram with perpendicular anisotropy, on the plane $K_z \times a$.}
\end{figure}

The phase diagram for the elliptic nanodots is also modified by the presence of perpendicular anisotropy. Its effect is illustrated in Fig. \ref{fig:EllipseDiagramKxa}, which shows the phase diagram for the spin configurations obtained by micromagnetic simulation for ellipses with thickness of 50 nm, as a function of the major semi-axis $a$, for different values of the perpendicular anisotropy. This diagram is more complex than that obtained for the disks with perpendicular anisotropy (Fig. \ref{fig:MicromagSimulRadius}). The letters in Figs. \ref{fig:EllipsePhaseDiagram} and \ref{fig:EllipseDiagramKxa} correspond to the spin configurations: A) in-plane magnetization (Fig. \ref{fig:EllipseConfigurations}A), B) perpendicular magnetization (Fig. \ref{fig:EllipseConfigurations}B), C) in-plane vortex (Fig. \ref{fig:EllipseConfigurations}C), D) double in-plane vortex (Fig. \ref{fig:EllipseConfigurations}D), E, F, G) types of lateral vortices (Fig. \ref{fig:EllipseLateral}). Some lateral vortex configurations also shown in the elliptic nanodot cross section are illustrated in Fig. \ref{fig:EllipseLateral}. They are: E) "two-domain" out-of-plane structure with one lateral vortex, F) modified in-plane vortex and G) "three-domain" out-of-plane structure with two lateral vortices of opposite polarization. The lateral vortices in E and G occur between two domains. Note that the contrast color of the top view of the ellipses is in the $z$ axis (perpendicular to the plane) and the cross section is shown with contrast in the $y$ axis, to make the vortex structures more visible; note also that although the ellipses in Fig. \ref{fig:EllipseLateral} are shown with the same size, they correspond to different semi-axes and anisotropies in the phase diagram of  Fig. \ref{fig:EllipseDiagramKxa}.

\section{Conclusions}

There are several ways of playing with the magnetic configurations of nanodots; the most obvious ones are to change their shape, for example, from circular to elliptical, or to vary their dimensions.
In this work we have explored a different way of accomplishing this: the introduction of a perpendicular anisotropy term. We have observed that this leads to important modifications in the phase diagrams for these nanostructures, as demonstrated through results obtained by micromagnetic simulation and analytical formulation. MFM measurements confirmed the trend of increasing vortex core diameter with increasing perpendicular anisotropy.

In this work we have studied the different magnetic configurations of circular and elliptical nanodots, presenting them using $h \times r$ phase diagrams obtained using micromagnetic simulation. In the case of circular nanodots, a phase diagram was also obtained using an analytical method; it agrees with the micromagnetic simulation. Measurements using the MFM technique show the same qualitative behavior in the dependence of the vortex core diameter with perpendicular magnetic anisotropy. The phase diagrams are also drawn for nanodots presenting a perpendicular anisotropy term, and exhibit important differences from the $K_z=0$ case: the region of the diagram corresponding to a magnetization perpendicular to the plane increases, the region of $M$ parallel to the plane is reduced, and more complex spin arrangements appear.

The results presented here on the elliptical nanodots reveal the complexity of their magnetic behavior. In the range of ellipse sizes studied, several configurations appear: in-plane and out-of-plane quasi-uniform states, one and two-vortex states, as well as configurations with lateral vortices. The latter structures obtained with the perpendicular anisotropy term are more complex and had not been investigated before; their detailed properties remain to be studied.

Investigations that allow the mapping of these different magnetic configurations are useful in designing experiments to study the basic properties of these novel magnetic structures, or tailoring them for technological applications, such as magnetic random access memories.

\begin{acknowledgments}

The authors would like to acknowledge the support of the Brazilian agencies CAPES, CNPq, FAPERJ and FAPEMIG. This work was partially funded by FONDECYT 11080246 and the program "Financiamiento Basal para Centros Cient\'{\i}ficos y Tecnol\'ogicos de Excelencia" CEDENNA FB0807, Chile.

\end{acknowledgments}



\begin{thebibliography}{31}%
\makeatletter
\providecommand \@ifxundefined [1]{%
 \@ifx{#1\undefined}
}%
\providecommand \@ifnum [1]{%
 \ifnum #1\expandafter \@firstoftwo
 \else \expandafter \@secondoftwo
 \fi
}%
\providecommand \@ifx [1]{%
 \ifx #1\expandafter \@firstoftwo
 \else \expandafter \@secondoftwo
 \fi
}%
\providecommand \natexlab [1]{#1}%
\providecommand \enquote  [1]{``#1''}%
\providecommand \bibnamefont  [1]{#1}%
\providecommand \bibfnamefont [1]{#1}%
\providecommand \citenamefont [1]{#1}%
\providecommand \href@noop [0]{\@secondoftwo}%
\providecommand \href [0]{\begingroup \@sanitize@url \@href}%
\providecommand \@href[1]{\@@startlink{#1}\@@href}%
\providecommand \@@href[1]{\endgroup#1\@@endlink}%
\providecommand \@sanitize@url [0]{\catcode `\\12\catcode `\$12\catcode
  `\&12\catcode `\#12\catcode `\^12\catcode `\_12\catcode `\%12\relax}%
\providecommand \@@startlink[1]{}%
\providecommand \@@endlink[0]{}%
\providecommand \url  [0]{\begingroup\@sanitize@url \@url }%
\providecommand \@url [1]{\endgroup\@href {#1}{\urlprefix }}%
\providecommand \urlprefix  [0]{URL }%
\providecommand \Eprint [0]{\href }%
\providecommand \doibase [0]{http://dx.doi.org/}%
\providecommand \selectlanguage [0]{\@gobble}%
\providecommand \bibinfo  [0]{\@secondoftwo}%
\providecommand \bibfield  [0]{\@secondoftwo}%
\providecommand \translation [1]{[#1]}%
\providecommand \BibitemOpen [0]{}%
\providecommand \bibitemStop [0]{}%
\providecommand \bibitemNoStop [0]{.\EOS\space}%
\providecommand \EOS [0]{\spacefactor3000\relax}%
\providecommand \BibitemShut  [1]{\csname bibitem#1\endcsname}%
\let\auto@bib@innerbib\@empty
\bibitem [{\citenamefont {Guimar{\~a}es}(2009)}]{Guimaraes:2009}%
  \BibitemOpen
  \bibfield  {author} {\bibinfo {author} {\bibfnamefont {A.~P.}\ \bibnamefont
  {Guimar{\~a}es}},\ }\href@noop {} {\emph {\bibinfo {title} {Principles of
  Nanomagnetism}}}\ (\bibinfo  {publisher} {Springer},\ \bibinfo {address}
  {Berlin},\ \bibinfo {year} {2009})\BibitemShut {NoStop}%
\bibitem [{\citenamefont {Hubert}\ and\ \citenamefont
  {Sch{\"a}fer}(1999)}]{Hubert:1999}%
  \BibitemOpen
  \bibfield  {author} {\bibinfo {author} {\bibfnamefont {A.}~\bibnamefont
  {Hubert}}\ and\ \bibinfo {author} {\bibfnamefont {R.}~\bibnamefont
  {Sch{\"a}fer}},\ }\href@noop {} {\emph {\bibinfo {title} {Magnetic {D}omains.
  {The} {A}nalysis of {M}agnetic {M}icrostructures}}}\ (\bibinfo  {publisher}
  {Springer},\ \bibinfo {address} {Berlin},\ \bibinfo {year}
  {1999})\BibitemShut {NoStop}%
\bibitem [{\citenamefont {Chien}\ \emph {et~al.}(2007)\citenamefont {Chien},
  \citenamefont {Zhu},\ and\ \citenamefont {Zhu}}]{Chien:2007}%
  \BibitemOpen
  \bibfield  {author} {\bibinfo {author} {\bibfnamefont {C.~L.}\ \bibnamefont
  {Chien}}, \bibinfo {author} {\bibfnamefont {F.~Q.}\ \bibnamefont {Zhu}}, \
  and\ \bibinfo {author} {\bibfnamefont {J.-G.}\ \bibnamefont {Zhu}},\
  }\href@noop {} {\bibfield  {journal} {\bibinfo  {journal} {Physics Today}\
  }\textbf {\bibinfo {volume} {60}},\ \bibinfo {pages} {40} (\bibinfo {year}
  {2007})}\BibitemShut {NoStop}%
\bibitem [{\citenamefont {Guslienko}\ \emph {et~al.}(2008)\citenamefont
  {Guslienko}, \citenamefont {Lee},\ and\ \citenamefont
  {Kim}}]{Guslienko:2008}%
  \BibitemOpen
  \bibfield  {author} {\bibinfo {author} {\bibfnamefont {K.~Y.}\ \bibnamefont
  {Guslienko}}, \bibinfo {author} {\bibfnamefont {K.-S.}\ \bibnamefont {Lee}},
  \ and\ \bibinfo {author} {\bibfnamefont {S.-K.}\ \bibnamefont {Kim}},\
  }\href@noop {} {\bibfield  {journal} {\bibinfo  {journal} {Phys. Rev. Lett.}\
  }\textbf {\bibinfo {volume} {100}},\ \bibinfo {pages} {027203} (\bibinfo
  {year} {2008})}\BibitemShut {NoStop}%
\bibitem [{\citenamefont {Shinjo}\ \emph {et~al.}(2000)\citenamefont {Shinjo},
  \citenamefont {Okuno}, \citenamefont {Hassdorf}, \citenamefont {Shigeto},\
  and\ \citenamefont {Ono}}]{Shinjo:2000}%
  \BibitemOpen
  \bibfield  {author} {\bibinfo {author} {\bibfnamefont {T.}~\bibnamefont
  {Shinjo}}, \bibinfo {author} {\bibfnamefont {T.}~\bibnamefont {Okuno}},
  \bibinfo {author} {\bibfnamefont {R.}~\bibnamefont {Hassdorf}}, \bibinfo
  {author} {\bibfnamefont {K.}~\bibnamefont {Shigeto}}, \ and\ \bibinfo
  {author} {\bibfnamefont {T.}~\bibnamefont {Ono}},\ }\href@noop {} {\bibfield
  {journal} {\bibinfo  {journal} {Science}\ }\textbf {\bibinfo {volume}
  {289}},\ \bibinfo {pages} {930} (\bibinfo {year} {2000})}\BibitemShut
  {NoStop}%
\bibitem [{\citenamefont {Choe}\ \emph {et~al.}(2004)\citenamefont {Choe},
  \citenamefont {Acremann}, \citenamefont {Scholl}, \citenamefont {Bauer},
  \citenamefont {Doran}, \citenamefont {Stohr},\ and\ \citenamefont
  {Padmore}}]{Choe:2004}%
  \BibitemOpen
  \bibfield  {author} {\bibinfo {author} {\bibfnamefont {S.-B.}\ \bibnamefont
  {Choe}}, \bibinfo {author} {\bibfnamefont {Y.}~\bibnamefont {Acremann}},
  \bibinfo {author} {\bibfnamefont {A.}~\bibnamefont {Scholl}}, \bibinfo
  {author} {\bibfnamefont {A.}~\bibnamefont {Bauer}}, \bibinfo {author}
  {\bibfnamefont {A.}~\bibnamefont {Doran}}, \bibinfo {author} {\bibfnamefont
  {J.}~\bibnamefont {Stohr}}, \ and\ \bibinfo {author} {\bibfnamefont {H.~A.}\
  \bibnamefont {Padmore}},\ }\href@noop {} {\bibfield  {journal} {\bibinfo
  {journal} {Science}\ }\textbf {\bibinfo {volume} {304}},\ \bibinfo {pages}
  {420} (\bibinfo {year} {2004})}\BibitemShut {NoStop}%
\bibitem [{\citenamefont {Cowburn}\ \emph {et~al.}(1999)\citenamefont
  {Cowburn}, \citenamefont {Koltsov}, \citenamefont {Adeyeye}, \citenamefont
  {Welland},\ and\ \citenamefont {Tricker}}]{Cowburn:1999}%
  \BibitemOpen
  \bibfield  {author} {\bibinfo {author} {\bibfnamefont {R.~P.}\ \bibnamefont
  {Cowburn}}, \bibinfo {author} {\bibfnamefont {D.~K.}\ \bibnamefont
  {Koltsov}}, \bibinfo {author} {\bibfnamefont {A.~O.}\ \bibnamefont
  {Adeyeye}}, \bibinfo {author} {\bibfnamefont {M.~E.}\ \bibnamefont
  {Welland}}, \ and\ \bibinfo {author} {\bibfnamefont {D.~M.}\ \bibnamefont
  {Tricker}},\ }\href@noop {} {\bibfield  {journal} {\bibinfo  {journal} {Phys.
  Rev. Lett.}\ }\textbf {\bibinfo {volume} {83}},\ \bibinfo {pages} {1042}
  (\bibinfo {year} {1999})}\BibitemShut {NoStop}%
\bibitem [{\citenamefont {d'Albuquerque~e Castro}\ \emph
  {et~al.}(2002)\citenamefont {d'Albuquerque~e Castro}, \citenamefont {Altbir},
  \citenamefont {Retamal},\ and\ \citenamefont {Vargas}}]{Albuquerque:2002}%
  \BibitemOpen
  \bibfield  {author} {\bibinfo {author} {\bibfnamefont {J.}~\bibnamefont
  {d'Albuquerque~e Castro}}, \bibinfo {author} {\bibfnamefont {D.}~\bibnamefont
  {Altbir}}, \bibinfo {author} {\bibfnamefont {J.~C.}\ \bibnamefont {Retamal}},
  \ and\ \bibinfo {author} {\bibfnamefont {P.}~\bibnamefont {Vargas}},\
  }\href@noop {} {\bibfield  {journal} {\bibinfo  {journal} {Phys. Rev. Lett.}\
  }\textbf {\bibinfo {volume} {88}},\ \bibinfo {pages} {237202} (\bibinfo
  {year} {2002})}\BibitemShut {NoStop}%
\bibitem [{\citenamefont {Scholz}\ \emph {et~al.}(2003)\citenamefont {Scholz},
  \citenamefont {Guslienko}, \citenamefont {Novosad}, \citenamefont {Suess},
  \citenamefont {Schrefl}, \citenamefont {Chantrell},\ and\ \citenamefont
  {Fidler}}]{Scholz:2003}%
  \BibitemOpen
  \bibfield  {author} {\bibinfo {author} {\bibfnamefont {W.}~\bibnamefont
  {Scholz}}, \bibinfo {author} {\bibfnamefont {K.}~\bibnamefont {Guslienko}},
  \bibinfo {author} {\bibfnamefont {V.}~\bibnamefont {Novosad}}, \bibinfo
  {author} {\bibfnamefont {D.}~\bibnamefont {Suess}}, \bibinfo {author}
  {\bibfnamefont {T.}~\bibnamefont {Schrefl}}, \bibinfo {author} {\bibfnamefont
  {R.}~\bibnamefont {Chantrell}}, \ and\ \bibinfo {author} {\bibfnamefont
  {J.}~\bibnamefont {Fidler}},\ }\href@noop {} {\bibfield  {journal} {\bibinfo
  {journal} {J. Magn. Magn. Mat.}\ }\textbf {\bibinfo {volume} {266}},\
  \bibinfo {pages} {155} (\bibinfo {year} {2003})}\BibitemShut {NoStop}%
\bibitem [{\citenamefont {Jubert}\ and\ \citenamefont
  {Allenspach}(2004)}]{Jubert:2004}%
  \BibitemOpen
  \bibfield  {author} {\bibinfo {author} {\bibfnamefont {P.-O.}\ \bibnamefont
  {Jubert}}\ and\ \bibinfo {author} {\bibfnamefont {R.}~\bibnamefont
  {Allenspach}},\ }\href@noop {} {\bibfield  {journal} {\bibinfo  {journal}
  {Phys. Rev. B}\ }\textbf {\bibinfo {volume} {70}},\ \bibinfo {pages} {144402}
  (\bibinfo {year} {2004})}\BibitemShut {NoStop}%
\bibitem [{\citenamefont {Landeros}\ \emph {et~al.}(2005)\citenamefont
  {Landeros}, \citenamefont {Escrig}, \citenamefont {Altbir}, \citenamefont
  {Laroze}, \citenamefont {d'Albuquerque~e Castro},\ and\ \citenamefont
  {Vargas}}]{Landeros:2005}%
  \BibitemOpen
  \bibfield  {author} {\bibinfo {author} {\bibfnamefont {P.}~\bibnamefont
  {Landeros}}, \bibinfo {author} {\bibfnamefont {J.}~\bibnamefont {Escrig}},
  \bibinfo {author} {\bibfnamefont {D.}~\bibnamefont {Altbir}}, \bibinfo
  {author} {\bibfnamefont {D.}~\bibnamefont {Laroze}}, \bibinfo {author}
  {\bibfnamefont {J.}~\bibnamefont {d'Albuquerque~e Castro}}, \ and\ \bibinfo
  {author} {\bibfnamefont {P.}~\bibnamefont {Vargas}},\ }\href@noop {}
  {\bibfield  {journal} {\bibinfo  {journal} {Phys. Rev. B}\ }\textbf {\bibinfo
  {volume} {71}},\ \bibinfo {pages} {094435} (\bibinfo {year}
  {2005})}\BibitemShut {NoStop}%
\bibitem [{\citenamefont {Kravchuk}\ \emph {et~al.}(2007)\citenamefont
  {Kravchuk}, \citenamefont {Sheka},\ and\ \citenamefont
  {Gaididei}}]{Kravchuk:2007}%
  \BibitemOpen
  \bibfield  {author} {\bibinfo {author} {\bibfnamefont {V.~P.}\ \bibnamefont
  {Kravchuk}}, \bibinfo {author} {\bibfnamefont {D.~D.}\ \bibnamefont {Sheka}},
  \ and\ \bibinfo {author} {\bibfnamefont {Y.~B.}\ \bibnamefont {Gaididei}},\
  }\href@noop {} {\bibfield  {journal} {\bibinfo  {journal} {J. Magn. Magn.
  Mat.}\ }\textbf {\bibinfo {volume} {310}},\ \bibinfo {pages} {116} (\bibinfo
  {year} {2007})}\BibitemShut {NoStop}%
\bibitem [{\citenamefont {Landeros}\ \emph {et~al.}(2008)\citenamefont
  {Landeros}, \citenamefont {Escrig},\ and\ \citenamefont
  {Altbir}}]{Landeros:2008}%
  \BibitemOpen
  \bibfield  {author} {\bibinfo {author} {\bibfnamefont {P.}~\bibnamefont
  {Landeros}}, \bibinfo {author} {\bibfnamefont {J.}~\bibnamefont {Escrig}}, \
  and\ \bibinfo {author} {\bibfnamefont {D.}~\bibnamefont {Altbir}},\ }in\
  \href@noop {} {\emph {\bibinfo {booktitle} {Electromagnetic, Magnetostatic,
  and Exchange-Interaction Vortices in Confined Magnetic Structures}}},\
  \bibinfo {editor} {edited by\ \bibinfo {editor} {\bibfnamefont {E.~O.}\
  \bibnamefont {Kamenetskii}}}\ (\bibinfo  {publisher} {Research Signpost},\
  \bibinfo {address} {Kerala},\ \bibinfo {year} {2008})\BibitemShut {NoStop}%
\bibitem [{\citenamefont {Thomson}\ \emph {et~al.}(2008)\citenamefont
  {Thomson}, \citenamefont {Abelman},\ and\ \citenamefont
  {Groenland}}]{Thomson:2008}%
  \BibitemOpen
  \bibfield  {author} {\bibinfo {author} {\bibfnamefont {T.}~\bibnamefont
  {Thomson}}, \bibinfo {author} {\bibfnamefont {L.}~\bibnamefont {Abelman}}, \
  and\ \bibinfo {author} {\bibfnamefont {H.}~\bibnamefont {Groenland}},\ }in\
  \href@noop {} {\emph {\bibinfo {booktitle} {Magnetic Nanostructures in Modern
  Technology}}},\ \bibinfo {editor} {edited by\ \bibinfo {editor}
  {\bibfnamefont {B.}~\bibnamefont {Azzerboni}}, \bibinfo {editor}
  {\bibfnamefont {G.}~\bibnamefont {Asti}}, \bibinfo {editor} {\bibfnamefont
  {L.}~\bibnamefont {Pareti}}, \ and\ \bibinfo {editor} {\bibfnamefont
  {M.}~\bibnamefont {Ghidini}}}\ (\bibinfo  {publisher} {Springer},\ \bibinfo
  {address} {Dordrecht},\ \bibinfo {year} {2008})\ pp.\ \bibinfo {pages}
  {237--306}\BibitemShut {NoStop}%
\bibitem [{\citenamefont {Kim}\ \emph {et~al.}(2008)\citenamefont {Kim},
  \citenamefont {Lee}, \citenamefont {Yu},\ and\ \citenamefont
  {Choi}}]{Kim:2008}%
  \BibitemOpen
  \bibfield  {author} {\bibinfo {author} {\bibfnamefont {S.}~\bibnamefont
  {Kim}}, \bibinfo {author} {\bibfnamefont {K.}~\bibnamefont {Lee}}, \bibinfo
  {author} {\bibfnamefont {Y.}~\bibnamefont {Yu}}, \ and\ \bibinfo {author}
  {\bibfnamefont {Y.}~\bibnamefont {Choi}},\ }\href@noop {} {\bibfield
  {journal} {\bibinfo  {journal} {Appl. Phys. Lett.}\ }\textbf {\bibinfo
  {volume} {92}},\ \bibinfo {pages} {022509} (\bibinfo {year}
  {2008})}\BibitemShut {NoStop}%
\bibitem [{\citenamefont {Bohlens}\ \emph {et~al.}(2008)\citenamefont
  {Bohlens}, \citenamefont {Kr{\"u}ger}, \citenamefont {Drews}, \citenamefont
  {Bolte}, \citenamefont {Meier},\ and\ \citenamefont
  {Pfannkuche}}]{Bohlens:2008}%
  \BibitemOpen
  \bibfield  {author} {\bibinfo {author} {\bibfnamefont {S.}~\bibnamefont
  {Bohlens}}, \bibinfo {author} {\bibfnamefont {B.}~\bibnamefont {Kr{\"u}ger}},
  \bibinfo {author} {\bibfnamefont {A.}~\bibnamefont {Drews}}, \bibinfo
  {author} {\bibfnamefont {M.}~\bibnamefont {Bolte}}, \bibinfo {author}
  {\bibfnamefont {G.}~\bibnamefont {Meier}}, \ and\ \bibinfo {author}
  {\bibfnamefont {D.}~\bibnamefont {Pfannkuche}},\ }\href@noop {} {\bibfield
  {journal} {\bibinfo  {journal} {Appl. Phys. Lett.}\ }\textbf {\bibinfo
  {volume} {93}},\ \bibinfo {pages} {142508} (\bibinfo {year}
  {2008})}\BibitemShut {NoStop}%
\bibitem [{\citenamefont {Lehndorff}\ \emph {et~al.}(2009)\citenamefont
  {Lehndorff}, \citenamefont {B\"urgler}, \citenamefont {Gliga}, \citenamefont
  {Hertel}, \citenamefont {Gr\"unberg}, \citenamefont {Schneider},\ and\
  \citenamefont {Celinski}}]{Lehndorff:2009}%
  \BibitemOpen
  \bibfield  {author} {\bibinfo {author} {\bibfnamefont {R.}~\bibnamefont
  {Lehndorff}}, \bibinfo {author} {\bibfnamefont {D.~E.}\ \bibnamefont
  {B\"urgler}}, \bibinfo {author} {\bibfnamefont {S.}~\bibnamefont {Gliga}},
  \bibinfo {author} {\bibfnamefont {R.}~\bibnamefont {Hertel}}, \bibinfo
  {author} {\bibfnamefont {P.}~\bibnamefont {Gr\"unberg}}, \bibinfo {author}
  {\bibfnamefont {C.~M.}\ \bibnamefont {Schneider}}, \ and\ \bibinfo {author}
  {\bibfnamefont {Z.}~\bibnamefont {Celinski}},\ }\href@noop {} {\bibfield
  {journal} {\bibinfo  {journal} {Phys. Rev. B}\ }\textbf {\bibinfo {volume}
  {80}},\ \bibinfo {pages} {054412} (\bibinfo {year} {2009})}\BibitemShut
  {NoStop}%
\bibitem [{\citenamefont {Houssameddine}\ \emph {et~al.}(2007)\citenamefont
  {Houssameddine}, \citenamefont {Ebels}, \citenamefont {Delaet}, \citenamefont
  {Rodmacq}, \citenamefont {Firastrau}, \citenamefont {Ponthenier},
  \citenamefont {Brunet}, \citenamefont {Thirion}, \citenamefont {Michel},
  \citenamefont {Prejbeanu-Buda}, \citenamefont {Cyrille}, \citenamefont
  {Redon},\ and\ \citenamefont {Dieny}}]{Houssameddine:2007}%
  \BibitemOpen
  \bibfield  {author} {\bibinfo {author} {\bibfnamefont {D.}~\bibnamefont
  {Houssameddine}}, \bibinfo {author} {\bibfnamefont {U.}~\bibnamefont
  {Ebels}}, \bibinfo {author} {\bibfnamefont {B.}~\bibnamefont {Delaet}},
  \bibinfo {author} {\bibfnamefont {B.}~\bibnamefont {Rodmacq}}, \bibinfo
  {author} {\bibfnamefont {I.}~\bibnamefont {Firastrau}}, \bibinfo {author}
  {\bibfnamefont {F.}~\bibnamefont {Ponthenier}}, \bibinfo {author}
  {\bibfnamefont {M.}~\bibnamefont {Brunet}}, \bibinfo {author} {\bibfnamefont
  {C.}~\bibnamefont {Thirion}}, \bibinfo {author} {\bibfnamefont {J.-P.}\
  \bibnamefont {Michel}}, \bibinfo {author} {\bibfnamefont {L.}~\bibnamefont
  {Prejbeanu-Buda}}, \bibinfo {author} {\bibfnamefont {M.-C.}\ \bibnamefont
  {Cyrille}}, \bibinfo {author} {\bibfnamefont {O.}~\bibnamefont {Redon}}, \
  and\ \bibinfo {author} {\bibfnamefont {B.}~\bibnamefont {Dieny}},\
  }\href@noop {} {\bibfield  {journal} {\bibinfo  {journal} {Nature Mater.}\
  }\textbf {\bibinfo {volume} {6}},\ \bibinfo {pages} {447} (\bibinfo {year}
  {2007})}\BibitemShut {NoStop}%
\bibitem [{\citenamefont {Berkov}\ and\ \citenamefont
  {Gorn}(2009)}]{Berkov:2009}%
  \BibitemOpen
  \bibfield  {author} {\bibinfo {author} {\bibfnamefont {D.~V.}\ \bibnamefont
  {Berkov}}\ and\ \bibinfo {author} {\bibfnamefont {N.~L.}\ \bibnamefont
  {Gorn}},\ }\href@noop {} {\bibfield  {journal} {\bibinfo  {journal} {Phys.
  Rev. B}\ }\textbf {\bibinfo {volume} {80}},\ \bibinfo {pages} {064409}
  (\bibinfo {year} {2009})}\BibitemShut {NoStop}%
\bibitem [{\citenamefont {Kim}\ \emph {et~al.}(2010)\citenamefont {Kim},
  \citenamefont {Rozhkova}, \citenamefont {Ulasov}, \citenamefont {Bader},
  \citenamefont {Rajh}, \citenamefont {Lesniak},\ and\ \citenamefont
  {Novosad}}]{Kim:2010}%
  \BibitemOpen
  \bibfield  {author} {\bibinfo {author} {\bibfnamefont {D.-H.}\ \bibnamefont
  {Kim}}, \bibinfo {author} {\bibfnamefont {E.~A.}\ \bibnamefont {Rozhkova}},
  \bibinfo {author} {\bibfnamefont {I.~V.}\ \bibnamefont {Ulasov}}, \bibinfo
  {author} {\bibfnamefont {S.~D.}\ \bibnamefont {Bader}}, \bibinfo {author}
  {\bibfnamefont {T.}~\bibnamefont {Rajh}}, \bibinfo {author} {\bibfnamefont
  {M.~S.}\ \bibnamefont {Lesniak}}, \ and\ \bibinfo {author} {\bibfnamefont
  {V.}~\bibnamefont {Novosad}},\ }\href@noop {} {\bibfield  {journal} {\bibinfo
   {journal} {Nature Mater.}\ }\textbf {\bibinfo {volume} {9}},\ \bibinfo
  {pages} {165} (\bibinfo {year} {2010})}\BibitemShut {NoStop}%
\bibitem [{\citenamefont {Ross}\ \emph {et~al.}(2002)\citenamefont {Ross},
  \citenamefont {Hwang}, \citenamefont {Shima}, \citenamefont {Cheng},
  \citenamefont {Farhoud}, \citenamefont {Savas}, \citenamefont {Smith},
  \citenamefont {Schwarzacher}, \citenamefont {Ross}, \citenamefont {Redjdal},\
  and\ \citenamefont {Humphrey}}]{Ross:2002}%
  \BibitemOpen
  \bibfield  {author} {\bibinfo {author} {\bibfnamefont {C.~A.}\ \bibnamefont
  {Ross}}, \bibinfo {author} {\bibfnamefont {M.}~\bibnamefont {Hwang}},
  \bibinfo {author} {\bibfnamefont {M.}~\bibnamefont {Shima}}, \bibinfo
  {author} {\bibfnamefont {J.~Y.}\ \bibnamefont {Cheng}}, \bibinfo {author}
  {\bibfnamefont {M.}~\bibnamefont {Farhoud}}, \bibinfo {author} {\bibfnamefont
  {T.~A.}\ \bibnamefont {Savas}}, \bibinfo {author} {\bibfnamefont {H.~I.}\
  \bibnamefont {Smith}}, \bibinfo {author} {\bibfnamefont {W.}~\bibnamefont
  {Schwarzacher}}, \bibinfo {author} {\bibfnamefont {F.~M.}\ \bibnamefont
  {Ross}}, \bibinfo {author} {\bibfnamefont {M.}~\bibnamefont {Redjdal}}, \
  and\ \bibinfo {author} {\bibfnamefont {F.~B.}\ \bibnamefont {Humphrey}},\
  }\href@noop {} {\bibfield  {journal} {\bibinfo  {journal} {Phys. Rev. B}\
  }\textbf {\bibinfo {volume} {65}},\ \bibinfo {pages} {144417} (\bibinfo
  {year} {2002})}\BibitemShut {NoStop}%
\bibitem [{\citenamefont {Chung}\ \emph {et~al.}(2010)\citenamefont {Chung},
  \citenamefont {McMichael}, \citenamefont {Pierce},\ and\ \citenamefont
  {Unguris}}]{Chung:2010}%
  \BibitemOpen
  \bibfield  {author} {\bibinfo {author} {\bibfnamefont {S.-H.}\ \bibnamefont
  {Chung}}, \bibinfo {author} {\bibfnamefont {R.~D.}\ \bibnamefont
  {McMichael}}, \bibinfo {author} {\bibfnamefont {D.~T.}\ \bibnamefont
  {Pierce}}, \ and\ \bibinfo {author} {\bibfnamefont {J.}~\bibnamefont
  {Unguris}},\ }\href@noop {} {\bibfield  {journal} {\bibinfo  {journal} {Phys.
  Rev. B}\ }\textbf {\bibinfo {volume} {81}},\ \bibinfo {pages} {024410}
  (\bibinfo {year} {2010})}\BibitemShut {NoStop}%
\bibitem [{\citenamefont {Metlov}\ and\ \citenamefont
  {Guslienko}(2002)}]{Metlov:2002}%
  \BibitemOpen
  \bibfield  {author} {\bibinfo {author} {\bibfnamefont {K.~L.}\ \bibnamefont
  {Metlov}}\ and\ \bibinfo {author} {\bibfnamefont {K.~Y.}\ \bibnamefont
  {Guslienko}},\ }\href@noop {} {\bibfield  {journal} {\bibinfo  {journal} {J.
  Magn. Magn. Mat.}\ }\textbf {\bibinfo {volume} {242-245}},\ \bibinfo {pages}
  {1015} (\bibinfo {year} {2002})}\BibitemShut {NoStop}%
\bibitem [{\citenamefont {Zhang}\ \emph {et~al.}(2008)\citenamefont {Zhang},
  \citenamefont {Singh}, \citenamefont {Bray-Ali},\ and\ \citenamefont
  {Haas}}]{Zhang:2008}%
  \BibitemOpen
  \bibfield  {author} {\bibinfo {author} {\bibfnamefont {W.}~\bibnamefont
  {Zhang}}, \bibinfo {author} {\bibfnamefont {R.}~\bibnamefont {Singh}},
  \bibinfo {author} {\bibfnamefont {N.}~\bibnamefont {Bray-Ali}}, \ and\
  \bibinfo {author} {\bibfnamefont {S.}~\bibnamefont {Haas}},\ }\href@noop {}
  {\bibfield  {journal} {\bibinfo  {journal} {Phys. Rev. B}\ }\textbf {\bibinfo
  {volume} {77}},\ \bibinfo {pages} {144428} (\bibinfo {year}
  {2008})}\BibitemShut {NoStop}%
\bibitem [{\citenamefont {Garcia}\ \emph {et~al.}(2010)\citenamefont {Garcia},
  \citenamefont {Westfahl}, \citenamefont {Schoenmaker}, \citenamefont
  {Carvalho}, \citenamefont {Santos}, \citenamefont {Pojar}, \citenamefont
  {Seabra}, \citenamefont {Belkhou}, \citenamefont {Bendounan}, \citenamefont
  {Novais},\ and\ \citenamefont {Guimar{\~a}es}}]{Garcia:2010}%
  \BibitemOpen
  \bibfield  {author} {\bibinfo {author} {\bibfnamefont {F.}~\bibnamefont
  {Garcia}}, \bibinfo {author} {\bibfnamefont {H.}~\bibnamefont {Westfahl}},
  \bibinfo {author} {\bibfnamefont {J.}~\bibnamefont {Schoenmaker}}, \bibinfo
  {author} {\bibfnamefont {E.~J.}\ \bibnamefont {Carvalho}}, \bibinfo {author}
  {\bibfnamefont {A.~D.}\ \bibnamefont {Santos}}, \bibinfo {author}
  {\bibfnamefont {M.}~\bibnamefont {Pojar}}, \bibinfo {author} {\bibfnamefont
  {A.~C.}\ \bibnamefont {Seabra}}, \bibinfo {author} {\bibfnamefont
  {R.}~\bibnamefont {Belkhou}}, \bibinfo {author} {\bibfnamefont
  {A.}~\bibnamefont {Bendounan}}, \bibinfo {author} {\bibfnamefont {E.~R.~P.}\
  \bibnamefont {Novais}}, \ and\ \bibinfo {author} {\bibfnamefont {A.~P.}\
  \bibnamefont {Guimar{\~a}es}},\ }\href@noop {} {\bibfield  {journal}
  {\bibinfo  {journal} {Appl. Phys. Lett.}\ }\textbf {\bibinfo {volume} {97}},\
  \bibinfo {eid} {022501} (\bibinfo {year} {2010})}\BibitemShut {NoStop}%
\bibitem [{\citenamefont {Novais}\ and\ \citenamefont
  {Guimar{\~a}es}(2009)}]{Novais:2009}%
  \BibitemOpen
  \bibfield  {author} {\bibinfo {author} {\bibfnamefont {E.~R.~P.}\
  \bibnamefont {Novais}}\ and\ \bibinfo {author} {\bibfnamefont {A.~P.}\
  \bibnamefont {Guimar{\~a}es}},\ }\href@noop {} {\bibfield  {journal}
  {\bibinfo  {journal} {arXiv:0909.5686v1}\ } (\bibinfo {year}
  {2009})}\BibitemShut {NoStop}%
\bibitem [{\citenamefont {Ha}\ \emph {et~al.}(2003)\citenamefont {Ha},
  \citenamefont {Hertel},\ and\ \citenamefont {Kirschner}}]{Ha:2003}%
  \BibitemOpen
  \bibfield  {author} {\bibinfo {author} {\bibfnamefont {J.~K.}\ \bibnamefont
  {Ha}}, \bibinfo {author} {\bibfnamefont {R.}~\bibnamefont {Hertel}}, \ and\
  \bibinfo {author} {\bibfnamefont {J.}~\bibnamefont {Kirschner}},\ }\href@noop
  {} {\bibfield  {journal} {\bibinfo  {journal} {Phys. Rev. B}\ }\textbf
  {\bibinfo {volume} {67}},\ \bibinfo {pages} {224432} (\bibinfo {year}
  {2003})}\BibitemShut {NoStop}%
\bibitem [{\citenamefont {Schneider}\ \emph {et~al.}(2003)\citenamefont
  {Schneider}, \citenamefont {Liszkowski}, \citenamefont {Rahm}, \citenamefont
  {Wegscheider}, \citenamefont {Weiss}, \citenamefont {Hoffmann},\ and\
  \citenamefont {Zweck}}]{Schneider:2003}%
  \BibitemOpen
  \bibfield  {author} {\bibinfo {author} {\bibfnamefont {M.}~\bibnamefont
  {Schneider}}, \bibinfo {author} {\bibfnamefont {J.}~\bibnamefont
  {Liszkowski}}, \bibinfo {author} {\bibfnamefont {M.}~\bibnamefont {Rahm}},
  \bibinfo {author} {\bibfnamefont {W.}~\bibnamefont {Wegscheider}}, \bibinfo
  {author} {\bibfnamefont {D.}~\bibnamefont {Weiss}}, \bibinfo {author}
  {\bibfnamefont {H.}~\bibnamefont {Hoffmann}}, \ and\ \bibinfo {author}
  {\bibfnamefont {J.}~\bibnamefont {Zweck}},\ }\href@noop {} {\bibfield
  {journal} {\bibinfo  {journal} {J. Phys. D: Appl. Phys.}\ }\textbf {\bibinfo
  {volume} {36}},\ \bibinfo {pages} {2239} (\bibinfo {year}
  {2003})}\BibitemShut {NoStop}%
\bibitem [{\citenamefont {Buchanan}\ \emph {et~al.}(2005)\citenamefont
  {Buchanan}, \citenamefont {Roy}, \citenamefont {Grimsditch}, \citenamefont
  {Fradin}, \citenamefont {Guslienko}, \citenamefont {Bader},\ and\
  \citenamefont {Novosad}}]{Buchanan:2005}%
  \BibitemOpen
  \bibfield  {author} {\bibinfo {author} {\bibfnamefont {K.~S.}\ \bibnamefont
  {Buchanan}}, \bibinfo {author} {\bibfnamefont {P.~E.}\ \bibnamefont {Roy}},
  \bibinfo {author} {\bibfnamefont {M.}~\bibnamefont {Grimsditch}}, \bibinfo
  {author} {\bibfnamefont {F.~Y.}\ \bibnamefont {Fradin}}, \bibinfo {author}
  {\bibfnamefont {K.~Y.}\ \bibnamefont {Guslienko}}, \bibinfo {author}
  {\bibfnamefont {S.~D.}\ \bibnamefont {Bader}}, \ and\ \bibinfo {author}
  {\bibfnamefont {V.}~\bibnamefont {Novosad}},\ }\href@noop {} {\bibfield
  {journal} {\bibinfo  {journal} {Nature Phys.}\ }\textbf {\bibinfo {volume}
  {1}},\ \bibinfo {pages} {172} (\bibinfo {year} {2005})}\BibitemShut {NoStop}%
\bibitem [{\citenamefont {Vavassori}\ \emph {et~al.}()\citenamefont
  {Vavassori}, \citenamefont {Zaluzec}, \citenamefont {Metlushko},
  \citenamefont {Novosad}, \citenamefont {Ilic},\ and\ \citenamefont
  {Grimsditch}}]{Vavassori:2004}%
  \BibitemOpen
  \bibfield  {author} {\bibinfo {author} {\bibfnamefont {P.}~\bibnamefont
  {Vavassori}}, \bibinfo {author} {\bibfnamefont {N.}~\bibnamefont {Zaluzec}},
  \bibinfo {author} {\bibfnamefont {V.}~\bibnamefont {Metlushko}}, \bibinfo
  {author} {\bibfnamefont {V.}~\bibnamefont {Novosad}}, \bibinfo {author}
  {\bibfnamefont {B.}~\bibnamefont {Ilic}}, \ and\ \bibinfo {author}
  {\bibfnamefont {M.}~\bibnamefont {Grimsditch}},\ }\href@noop {} {\bibfield
  {journal} {\bibinfo  {journal} {Phys. Rev. B}\ }\textbf {\bibinfo {volume}
  {69}}}\BibitemShut {NoStop}%
\bibitem [{\citenamefont {Okuno}\ \emph {et~al.}(2004)\citenamefont {Okuno},
  \citenamefont {Mibu},\ and\ \citenamefont {Shinjo}}]{Okuno:2004}%
  \BibitemOpen
  \bibfield  {author} {\bibinfo {author} {\bibfnamefont {T.}~\bibnamefont
  {Okuno}}, \bibinfo {author} {\bibfnamefont {K.}~\bibnamefont {Mibu}}, \ and\
  \bibinfo {author} {\bibfnamefont {T.}~\bibnamefont {Shinjo}},\ }\href@noop {}
  {\bibfield  {journal} {\bibinfo  {journal} {J. Appl. Phys.}\ }\textbf
  {\bibinfo {volume} {95}},\ \bibinfo {pages} {3612} (\bibinfo {year}
  {2004})}\BibitemShut {NoStop}%
\end{thebibliography}


%


\end{document}